\documentclass[aps,prc,twocolumn,showpacs]{revtex4}
\usepackage{graphicx,amsfonts,amssymb,amsbsy}
\usepackage{pstricks}

\begin{document}

\title{Surface tension and curvature energy of quark matter in the NJL model}
\author{G. Lugones$^1$, A. G. Grunfeld$^{2,3}$ and M. Al Ajmi$^4$}
 \affiliation{ $^1$ Universidade Federal do ABC, Rua Santa Ad\'elia, 166, 09210-170, Santo Andr\'e, Brazil\\
 $^2$ CONICET, Rivadavia 1917, (1033) Buenos Aires, Argentina.\\
 $^3$ Departamento de F\'\i sica, Comisi\'on Nacional de
 Energ\'{\i}a At\'omica, (1429) Buenos Aires, Argentina \\
  $^4$ Department of Physics, Sultan Qaboos University, P.O.Box: 36 Al-Khode 123 Muscat, Sultanate of Oman.
}

\begin{abstract}
In this paper we study the surface tension and the curvature energy of three-flavor quark matter in equilibrium under weak interactions within the Nambu-Jona-Lasinio model. We include the effect of color superconductivity and describe finite size effects within the multiple reflection expansion (MRE) framework. Our calculations result in large values of the surface tension which disfavor the formation of mixed phases at the hadron-quark inter-phase inside a hybrid star. 
\end{abstract}


\maketitle
\section{Introduction}

The study of the surface tension of deconfined quark matter has attracted much attention recently \cite{Buballa2013,Pinto2012,Wen2010,Palhares2010} because a detailed knowledge of it may contribute to a better comprehension of the physics of compact star interiors. In fact, surface tension plays a crucial role in quark matter nucleation during the formation of compact stellar objects, because it determines the nucleation rate and the associated critical size of the nucleated drops \cite{doCarmo2013,Lugones2011}. It is also determinant in the formation of mixed phases at the core of hybrid stars which may arise only if the surface tension is smaller than a critical value of the order of tens of  MeV / fm$^2$  \cite{Voskresensky2003,Tatsumi2003,Maruyama2007,Endo2011}. Also,  surface tension affects decisively the properties of the most external layers of a strange star which may fragment into a charge-separated mixture, involving positively-charged strangelets immersed in a negatively charged sea of electrons, presumably forming a crystalline solid crust \cite{Jaikumar2006}. This would happen below a critical surface tension which is typically of the order of a few MeV/fm$^2$ \cite{Alford2006}.

However, in spite of its key role for compact star physics, the surface tension is still poorly known for quark matter. Early calculations by Berger and Jaffe gave rather low values for the surface tension, below $5  \, \mathrm{MeV/fm^2}$ \cite{Berger1987}. However, larger values within  $10 - 50 \, \mathrm{MeV/fm^2}$ were used in other works about quark matter droplets in neutron stars \cite{Heiselberg1993,Iida1998}. More recently,  values around $\approx 30 \, \mathrm{MeV/fm^2}$ have been adopted for studying the effect of quark matter nucleation on the evolution of protoneutron stars \cite{Bombaci2007,Bombaci2009}. However,
much larger values have also been obtained in the literature. Estimates given in  Ref. \cite{Voskresensky2003} give values in the range $50 - 150 \, \mathrm{MeV/fm^2}$ and 
values around  $\sim 300 \, \mathrm{MeV/fm^2}$ were suggested on the basis of dimensional analysis of the minimal interface between a color-flavor locked phase and nuclear matter \cite{Alford2001}.

In this paper we study the surface tension and the curvature energy of three-flavour quark matter in equilibrium under weak interactions within the Nambu-Jona-Lasinio (NJL) model. We include the effect of color superconductivity and describe finite size effects within the multiple reflection expansion (MRE) framework \cite{Balian1970,Madsen-drop,Kiriyama1,Kiriyama2}. We shall consider only the 2SC phase, but we shall see that our conclusions are quite general  and the results for other superconducting phases can be easily foreseen within the present model.  Our calculations result in large values of the surface tension which disfavor the formation of mixed phases at hybrid star cores.  

The article is organized as follows: in Sect. II we present the quark matter equations of state without finite size effects. Then, in Sect. III we introduce the MRE formalism for the finite size effects. Finally, in sect. IV we present our results and  conclusions.

%----------------------------------------------------------------------------------------
\section{The model in the bulk}
%----------------------------------------------------------------------------------------

In the present work we start from an $SU(3)_f$ NJL effective model which also includes color superconducting quark-quark interactions. The corresponding Lagrangian is given by 
\begin{eqnarray}
{\cal L} &=& \bar \psi \left(i \rlap/\partial - \hat m \right) \psi \nonumber \\
& + &  G \sum_{a=0}^8 \left[ \left( \bar \psi \ \tau_a \ \psi \right)^2 + \left( \bar \psi \ i \gamma_5 \tau_a \ \psi \right)^2 \right]
\label{lagrangian} \\
& + &  2H \!\! \sum_{A,A'=2,5,7} \left[ \left( \bar \psi \ i \gamma_5 \tau_A \lambda_{A'} \ \psi_C \right) \left( \bar \psi_C \ i \gamma_5 \tau_A \lambda_{A'} \ \psi \right) \right] \nonumber
\end{eqnarray}
where $\hat m=\mathrm{diag}(m_u,m_d,m_s)$ is the current mass
matrix in flavour space. In what follows we will work in the
isospin symmetric limit $m_u=m_d=m$, and for simplicity we do not include flavour mixing effects. The matrices $\tau_i$ and
$\lambda_i$ with $i=1,..,8$ are the Gell-Mann matrices
corresponding to the flavour and color groups respectively, and
$\tau_0 = \sqrt{2/3}\ 1_f$. In addition, in Eq. (\ref{lagrangian}) we have used the charge conjugate spinors
$\psi_C = C \ \bar \psi^T$ and $\bar
\psi_C = \psi^T C$, where $\bar \psi = \psi^\dagger \gamma^0$ is
the Dirac conjugate spinor and $C=i\gamma^2 \gamma^0$.

The next step is obtaining the grand canonical thermodynamic potential at
finite temperature $T$ and chemical potentials $\mu_{fc}$, where
$f$ and $c$ stand for flavour and color respectively. Then, we are able to calculate the relevant thermodynamic quantities. Note that first we will show the thermodynamic potential for the bulk system, and later on we will derive the effective potential for the finite size effects.
For that purpose, starting from Eq. (\ref{lagrangian}), it is convenient to perform a standard bosonization of the theory.
Thus, we introduce the bosonic fields $\sigma_a$, $\pi_a$ and $\Delta_A$ corresponding
to the sigma and pion mesons, and scalar diquark fields, respectively; and integrate out the quark fields.
In what follows we will work within the mean field approximation (MFA), in
which these bosonic fields are expanded around their vacuum
expectation values and the corresponding fluctuations are neglected. Since the mean field
values of the pion fields vanish due to symmetry reasons,
in what follows we will only consider those of the sigma and diquark fields. Then, $\hat \sigma = \sigma_a \tau_a =
\textrm{diag}(\sigma_u,\sigma_d,\sigma_s)$ and $\pi_a=0$. Regarding the diquark mean field, in the present work, we will assume that in the density region of interest only the 2SC phase might be relevant. Moreover, due to the color symmetry, one can rotate in color space to fix $\Delta_5 = \Delta_7 = 0$, $\Delta_2 = \Delta$. Finally, in the
framework of the Matsubara and Nambu-Gorkov formalism we obtain the following MFA thermodynamic potential
$\Omega_{MFA}(T,\mu_{fc},\sigma_u,\sigma_d,\sigma_s,|\Delta|)$
per unit volume (further calculation details can be found in
Refs. \cite{Huang:2002zd,Ruester:2005jc,Blaschke:2005uj,Abuki2005,Abuki2006,Ciminale,Hatsuda:1994pi,Buballa2005})
%}} 
%
\begin{eqnarray}
\frac{\Omega_{MFA}}{V}  & = &  2 \int_0^\Lambda \frac{k^2dk }{2  \; \pi^2} \sum_{i=1}^9  \omega(x_i,y_i)  \nonumber \\  & + & \frac{1}{4 G} (\sigma_u^2 + \sigma_d^2 + \sigma_s^2) + \frac{|\Delta|^2}{2H},
\end{eqnarray}
where $\Lambda$ is the cut-off of the model and $\omega(x,y)$ is defined by
\begin{eqnarray}
\omega(x,y)& = & - x - T \ln[1+e^{-(x-y)/T}] \nonumber \\
           &   &  -  T \ln[1+e^{-(x+y)/T}]  ,
\label{omeguinha}
\end{eqnarray}	
with 
\begin{eqnarray}
x_{1,2} = E ,  \quad  x_{3,4,5} = E_s,  \\ 
x_{6,7} = ([ E \pm  (\mu_{ur} + \mu_{dg})/2 ]^2 + \Delta^2 )^{1/2},  \\ 
x_{8,9} = ([ E \pm  (\mu_{ug} + \mu_{dr})/2 ]^2 + \Delta^2)^{1/2} \ , \\ 
y_1     =   \mu_{ub}  , \quad  y_2 = \mu_{db} , \quad   y_{3} = \mu_{sr}, \\
y_{4}   = \mu_{sg} , \quad  y_{5} = \mu_{sb} , \\ 
y_{6,7} = (\mu_{ur}-\mu_{dg})/2 , \quad  y_{8,9} = (\mu_{ug} - \mu_{dr})/2 . 
\end{eqnarray}
In the above expressions $E = \sqrt{ k^2+ M^2}$ and $E_s=\sqrt{  k ^2 + M_s^2}$,
with $M_f = m_f + \sigma_f$. As we are working in the isospin limit, then
$\sigma_u = \sigma_d = \sigma$ which gives $M_u = M_d =
M$. In principle one has nine different quark chemical potentials,
corresponding to the three quark flavours (u, d and s) and three quark colors (r, g,
and b). Nevertheless, as discussed above, with our particular election of the orientation of the gap $\Delta$ in the color space, 
there is a residual color symmetry (between red and green colors). Moreover, if we require the system to be in
chemical equilibrium, it can be seen that all chemical potentials
are not independent from each other, as it will be discussed in next section.

The total thermodynamic potential is obtained by adding to $\Omega_{MFA}$ the
contribution of the {{electrons}} and a vacuum constant. Namely,
\begin{equation}
\Omega = \Omega_{MFA} + \Omega_e  - \Omega_\textrm{\tiny vac} \label{QMP}
\end{equation}
where $\Omega_e$ is the thermodynamic potential of the electrons. For them we use the expression corresponding to a free gas of
ultra-relativistic fermions
\begin{equation}
\frac{\Omega_e(T,\mu_{e}) }{V} =  - 2 \left( \frac{\mu_e^4}{24 \pi^2} +
\frac{\mu_e^2T^2}{12} + \frac{7\pi^2T^4}{360} \right).\nonumber
\end{equation}
It is important to notice that in Eq. (\ref{QMP}) we have subtracted the constant $\Omega_\textrm{\tiny vac} \equiv - P_\textrm{\tiny vac} V$ in order to have a vanishing pressure at vanishing temperature and chemical potentials.
However, this conventional prescription is merely an arbitrary way to uniquely determine the EoS of the NJL model without any further assumptions \cite{Lenzi2012}. In the MIT bag model for instance, the pressure in the vacuum is non-vanishing. In view of this, $\Omega_{vac}$ is taken as a free parameter  in Ref. \cite{Lenzi2012}, having in mind that tuning this constant is an easy way to control the splitting between the chiral restoration density and the deconfinement  density. Nevertheless, we shall see below that $\Omega_{vac}$ doesn't have a direct influence on the values of the surface tension and the curvature energy.

\vspace{0.5 cm}

%........................................................................................
\section{Finite size effects}
%........................................................................................
%
\subsection{MRE formalism}
Now we are ready to introduce the effects of finite size in the thermodynamic potential. For doing so we consider the multiple reflection expansion formalism (see Refs. \cite{Balian1970, Madsen-drop,Kiriyama1,Kiriyama2} and references therein) which consists in modifying the density of states for the case of a finite spherical droplet as follows 
\begin{equation}
\rho_{MRE}(k,m_f,R) = 1 + \frac{6\pi^2}{kR} f_S + \frac{12\pi^2}{(kR)^2} f_C
 \end{equation}
where the surface contribution to the density of states is
\begin{equation}
f_S  = - \frac{1}{8 \pi} \left(1 -\frac{2}{\pi} \arctan \frac{k}{m_f} \right), 
\end{equation}
and the curvature contribution is given in the Madsen ansatz \cite{Madsen-drop}
\begin{equation}
f_C  =  \frac{1}{12 \pi^2} \left[1 -\frac{3k}{2m_f} \left(\frac{\pi}{2} - \arctan \frac{k}{m_f} \right)\right] 
\end{equation}
to take into account the finite quark mass contribution.

The density of states of MRE for massive quarks is reduced compared with the bulk one, and for a 
range of small momentum becomes negative. This non-physical negative values are removed
by introducing an infrared (IR) cutoff in momentum space \cite{Kiriyama2}. Thus,  we have to perform the following replacement in order to obtain the thermodynamic quantities
\begin{equation}
\int_0^{\Lambda} {{\cdots}} \frac{k^2 \, dk}{2 \pi^2}  \longrightarrow
\int_{\Lambda_{IR}}^\Lambda {{\cdots}} \frac{k^2 \, dk}{2 \pi^2} \rho_{MRE}.
\label{MRE}
\end{equation}
The IR cut-off $\Lambda_{IR}$ is the largest solution of the
equation $\rho_{MRE} (k)= 0$ with respect to the momentum $k$. 

After the above replacement, the full thermodynamic potential for finite size spherical droplets reads:
\begin{eqnarray}
\frac{\Omega_{{MRE}}}{V} & = & 2
\int_{{{\Lambda_{IR}}}}^\Lambda \frac{k^2dk }{2 \; \pi^2}
\rho_{MRE} \sum_{i=1}^9  \omega(x_i,y_i)    \nonumber \\
   & &  + \frac{1}{4G}(\sigma_u^2+\sigma_d^2+\sigma_s^2)  + \frac{|\Delta|^2}{2H}  \nonumber \\
   & &  - P_e  + P_\textrm{\tiny vac} .
\label{fullomega}
\end{eqnarray}
Multiplying on both sides of the last equation by the volume of
the quark matter drop and rearranging terms we arrive to the
following form for $\Omega_{MRE}$ 
\begin{equation}
\Omega_{{MRE}} = -P V + \alpha S + \gamma C ,
\label{eq17}
\end{equation}
where the pressure $P$, the surface tension and the curvature energy density, are defined as \cite{Lugones2011}
\begin{eqnarray}
P   \equiv  - \frac{\partial \Omega_{{MRE}}}{ \partial V }
\bigg|_{T, \mu, S, C}  & = & - 2 \int_{{{\Lambda_{IR}}}}^\Lambda
\frac{k^2dk }{2  \; \pi^2} \sum_{i=1}^9 \omega(x_i,y_i)  \nonumber\\
 &  &- \frac{1}{4G}(\sigma_u^2+\sigma_d^2+\sigma_s^2)  - \frac{|\Delta|^2}{2H}  \nonumber\\
 & &  + P_e  - P_\textrm{\tiny vac} ,
\end{eqnarray}
\begin{equation}
\alpha \equiv  \frac{\partial \Omega_{{MRE}}}{ \partial S }
\bigg|_{T, \mu, V, C}  = 2 \int_{{{\Lambda_{IR}}}}^\Lambda k \;
dk \; f_S \sum_{i=1}^9 \omega(x_i,y_i) ,
\label{surfacetension}
\end{equation}
and
\begin{equation}
\gamma \equiv  \frac{\partial \Omega_{{MRE}}}{ \partial C } \bigg|_{T, \mu, V, S}  = 2 \int_{{{\Lambda_{IR}}}}^\Lambda  dk \; f_C \sum_{i=1}^9 \omega(x_i,y_i).
\label{curvatureenergy}
\end{equation}
respectively.  {{We}} are considering a spherical drop, i.e.  the area is $S= 4\pi R^2$  and the curvature is $C=8\pi R$.

As mentioned above, once we have the grand thermodynamic potential $\Omega_{MRE}$ then we can obtain the relevant thermodynamic quantities. We can readily obtain the number density of quarks of each flavor and color 
\begin{eqnarray}
n_{fc} \equiv  \frac{1}{V} \frac{\partial \Omega_{{MRE}}}{ \partial \mu_{fc}  } 
\end{eqnarray}
and the number density of  $e^-$  (assumed to be massless)
\begin{eqnarray}
n_{e} \equiv  \frac{1}{V} \frac{\partial \Omega_{{MRE}} }{\partial \mu_e} .
\end{eqnarray}

Then, the corresponding number densities of each flavor, $n_f$, and of each color, $n_c$, in the quark phase are given by
$n_f = \sum_{c} n_{fc}$  and  $n_c = \sum_{f} n_{fc}$ respectively.  The baryon number density reads $n_B = \frac{1}{3}
\sum_{fc} n_{fc} = (n_u + n_d + n_s)/3$. 

%........................................................................................
\subsection{Neutrality conditions and $\beta$ decay}
%........................................................................................

In order to derive the EOS from the above formalism it
is necessary to impose a suitable number of conditions on the
variables  $\{\mu_{fc}\}, \mu_e, \sigma, \sigma_s$ and
$\Delta$. The first three of these conditions arise from the fact
that the thermodynamically consistent solutions correspond to the
stationary points of $\Omega_{{MRE}}$ with respect to
$\sigma$, $\sigma_s$, and $\Delta$. Then, we have
\begin{eqnarray}
\frac{\partial\Omega_{{MRE}}}{\partial\sigma} =0, \quad
\frac{\partial\Omega_{{MRE}}}{\partial\sigma_s} =0, \quad
\frac{\partial\Omega_{{MRE}}}{\partial|\Delta|}=0.
\label{gapeq}
\end{eqnarray}

For the remaining conditions one must specify the physical
situation in which one is interested in. In this work we are interested in the study of finite size color-superconducting droplets in $\beta$-equilibrium that may form  e.g. within the mixed phase of a hybrid star. In such a case, chemical
equilibrium is maintained by weak interactions among quarks, e.g.
$d \leftrightarrow u + e^- + \bar{\nu}_e$, $s \leftrightarrow u +
e^- + \bar{\nu}_e$, $u + d \leftrightarrow u + s$. Here we consider the situation of no neutrino trapping. 
Then, the lepton number is not conserved and we have four
independent conserved charges, namely the electric charge $n_Q =
\frac{2}{3} n_u - \frac{1}{3} n_d - \frac{1}{3} n_s - n_e$ and the
three color charges $n_u$, $n_d$ and $n_s$. It is more convenient to use 
the linear combinations $n = n_r + n_g + n_b$,
$n_3 = n_r - n_g$ and $n_8 = \frac{1}{\sqrt{3}}  (n_r + n_g - 2
n_b)$, where $n = 3 n_B$ (the total quark number density)
and $n_3$ and $n_8$ are related with color asymmetries. Thus, conserved charges $\{n_j\} = \{n, n_3, n_8,
n_Q\}$ are related to four independent chemical potentials
$\{\mu_j\} =\{\mu, \mu_3, \mu_8, \mu_Q\}$ such that $n_j = - {
\partial \Omega_{MRE} }/{\partial \mu_j}$. The individual quark chemical
potentials $\mu_{fc}$ are given by
\begin{eqnarray}
\mu_{fc} &=& \mu + \mu_Q \left[ \frac{1}{2} (\tau_3)_{ff} +
\frac{1}{2\sqrt{3}} (\tau_8)_{ff} \right]
\nonumber \\     & &  \;
+ \mu_3 (\lambda_3)_{cc}  + \mu_8 (\lambda_8)_{cc}.
\end{eqnarray}
where, as before, $\tau_i$  and $\lambda_i$ are the Gell-Mann
matrices in flavor and color space respectively. 

From the $\beta$-equilibrium conditions we have 
\begin{eqnarray}
\mu_{dc} = \mu_{sc} = \mu_{uc} + \mu_e
\label{beta}
\end{eqnarray}
for all colors $c$. Then, the electron chemical potential is $\mu_e = - \mu_Q$. 
Finally, the rest of the conditions we need to impose for electrically and color neutral matter are:
\begin{eqnarray}
n_Q \equiv - \frac{\partial \Omega }{\partial \mu_Q} = 0, \quad \nonumber \\
%n_3 \equiv - \frac{\partial \Omega }{\partial \mu_3} = 0, \quad  \nonumber \\
n_8 \equiv - \frac{\partial \Omega }{\partial \mu_8} = 0. 
\label{colden}
\end{eqnarray}
{{(Note: remember that we choose a particular orientation of {{the}} gap in the color space, which introduces the $r-g$ symmetry. Thus, we trivially satisfy $n_3 = 0$, and  then $\mu_3$ is automatically equal to zero.) }}
 
In summary, in the case of neutron star quark matter without neutrino trapping, for each
value of $\mu$ (or $\mu_B$) and $T$ one can find the values of $\sigma, \sigma_s, \Delta , \mu_e$ and $\mu_8$
by solving Eqs. {{(\ref{gapeq}) and (\ref{colden}) {{supplemented}} by (\ref{beta})}}. 
This allows us to obtain the quark matter EoS in the thermodynamic
region we are interested in and we can evaluate the curvature energy and surface tension of the color superconducting droplets.

It is important to remark that, in general, when we numerically solve the set of equations related with all the conditions discussed above, there might be
regions for which there is more than one solution for
each value of T and $\mu$. To choose the stable solution among all of them, we require it to be an overall minimum of the thermodynamic potential.

%----------------------------------------------------------------------------------------
\subsection{Parametrization}
%----------------------------------------------------------------------------------------

The set of parameters we use in the present work is the following (those in Ref. \cite{Hatsuda:1994pi} but without 't Hooft interactions), $m_{u,d}$ = 5.5 {{MeV}},  $m_s$ = 112.0 MeV,  $\Lambda$ = 602.3 {{MeV}} and
$G\Lambda^2$ = 4.638. Moreover, we considered the ratio $H/G = 3/4$ obtained from Fierz transformations of the one-gluon exchange interactions.

As we impose vanishing pressure at vanishing temperature and chemical potentials for Fermi momentum $k_F \rightarrow 0$ (when $R \rightarrow \infty$) \cite{Kiriyama1}, thus, $P_0$ is the same as in the bulk case. For the set of parameters we used, we found $\Omega_\textrm{\tiny vac}/ V =  -P_\textrm{\tiny vac} = -4301$ MeV/fm$^3$). However, we emphasize that the values of the surface tension $\alpha$ and the curvature energy $\gamma$ are not changed by the choice of $P_{vac}$.

As we previously mentioned, the value of $\Lambda_{IR}$ is the largest root when solving $\rho_{MRE} = 0$ with respect to $k$, depending on $m_f$ and R. We find that these solutions can
be fitted through the following rule,
\begin{equation}
\Lambda_{IR} = a \, R^b \label{LIR}
\end{equation}
with $R$ in fm and $\Lambda_{IR}$ in MeV. The coefficients for $m_u = 5.5$ MeV are $a = 135.45$ and $b = -0.85$. For $m_s = 112$
MeV we have  $a= 228.60$ and $b= -0.87$.

\vspace{0.5cm}

%-------------------------------------------------------------------------
\section{Results and Conclusions}
%-------------------------------------------------------------------------

\begin{figure*}
\includegraphics[scale=0.33]{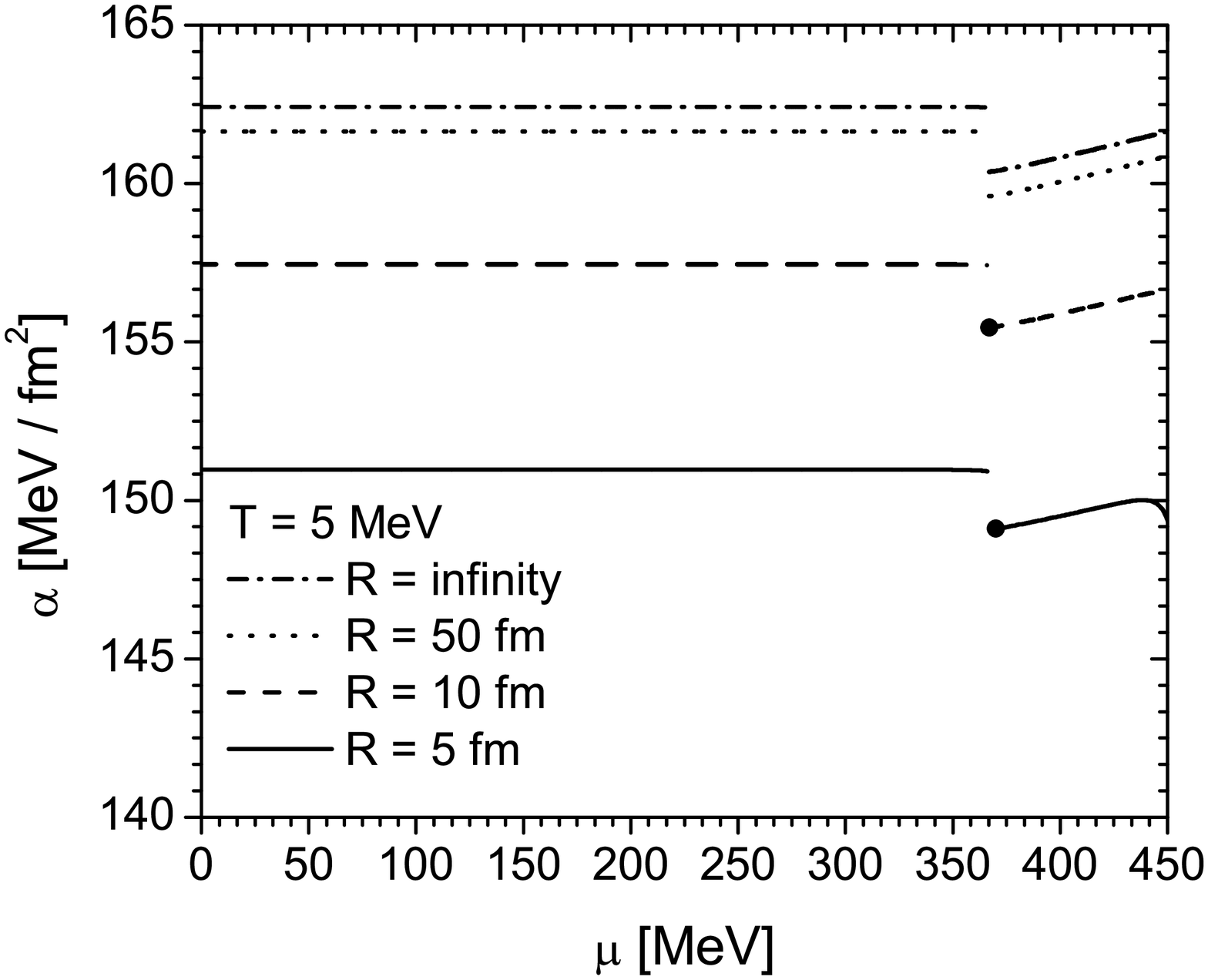}
\includegraphics[scale=0.33]{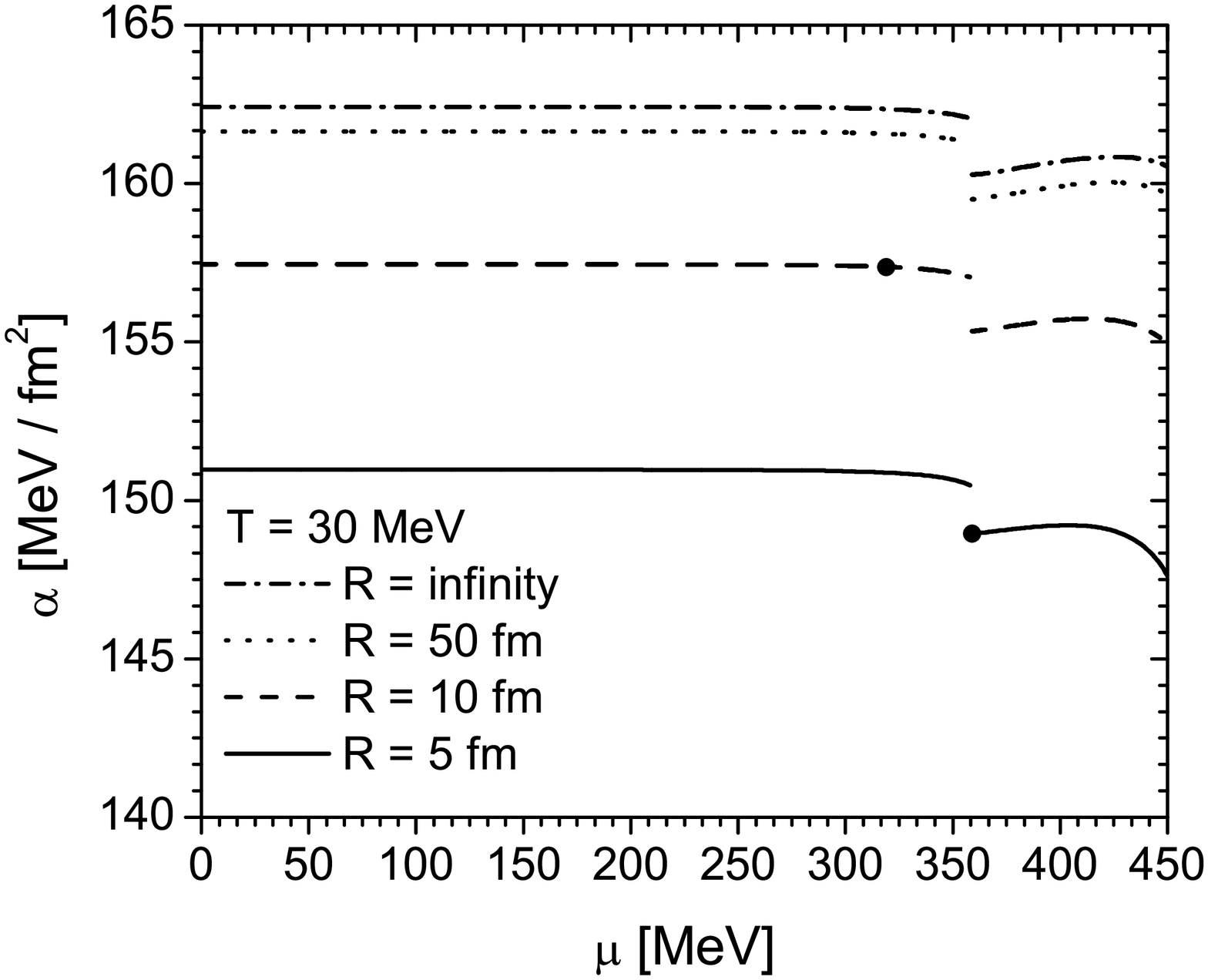}
\includegraphics[scale=0.33]{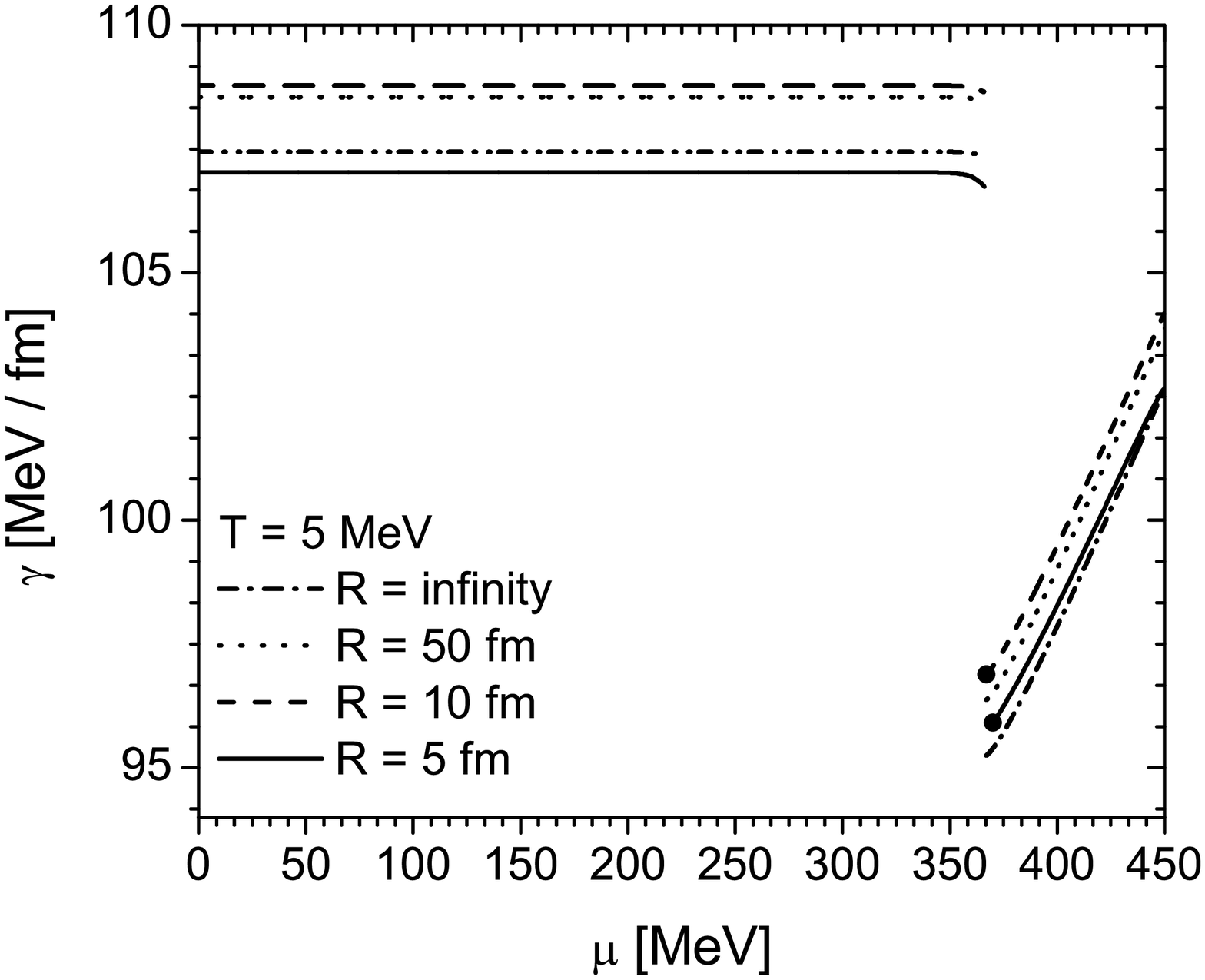}
\includegraphics[scale=0.33]{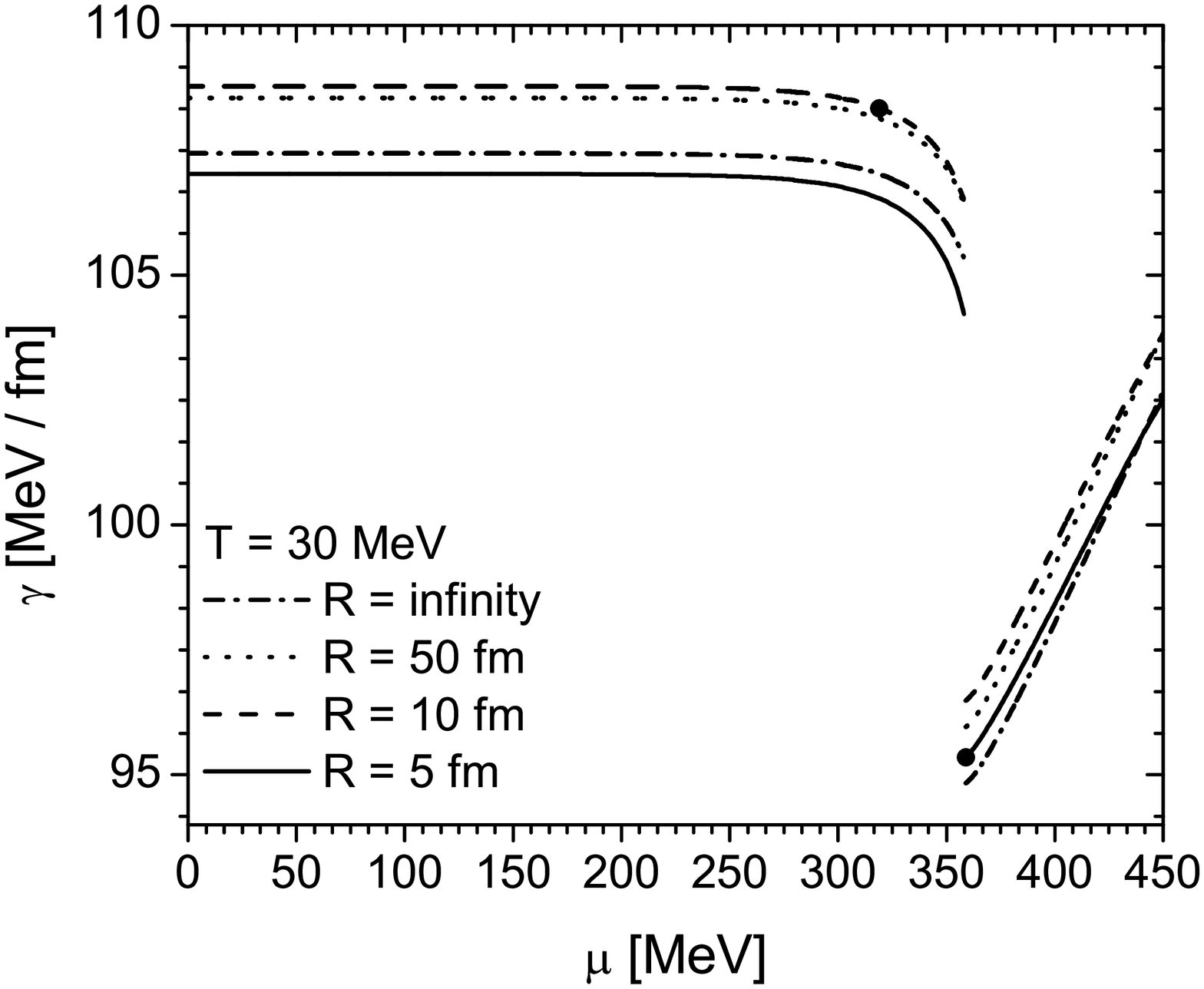}
\caption{Surface tension $\alpha$ and curvature energy $\gamma$ for different radii and temperatures. In all figures, the left branches correspond to the solutions in which the chiral symmetry broken phase is more favorable and the right curves after the discontinuity, correspond to the 2SC solutions.}
\label{fig1}
\end{figure*}

In Fig. \ref{fig1} we show our results for the surface tension $\alpha$ and curvature energy $\gamma$ of the color superconducting droplets as a function of the quark chemical potential, for two different temperatures and {{four}} different radii. The temperatures 30 MeV and 5 MeV are representative, respectively,  of the conditions prevailing at the beginning and at the end of the cooling/deleptonization phase of protoneutron star evolution. We have checked that the results for $T= 5$ MeV are almost indistinguishable of those for zero temperature, thus, in practice they also represent old and cold neutron stars. We show results for drops with radii ranging from very small values of  5 fm, which have a large energy cost due to surface and curvature effects, to the bulk limit of $R = \infty$.
For given values of $R$, $T$ and $\mu$ there may exist more than one solution of the equations. If more than one solution is found, the one that minimizes the thermodynamic potential is chosen. As explained in the figure caption, the left branches correspond to the chiral symmetry broken phase and the right curves after the discontinuity to the 2SC phase.
For the curves presenting negative pressures we have introduced a dot indicating the zero pressure point. The part of the curve to the left of the dot corresponds to $P \leq 0$. Note that, as previously mentioned, we subtracted $\Omega_{vac}$ to have vanishing pressure at vanishing $T$ and $\mu$ for $R = \infty$. However, $\Omega_{vac}$ can be taken as a free parameter as in Ref. \cite{Lenzi2012}. In this case,  the dot in the figures can move along the curves. However, it is clear from Eqs. (\ref{surfacetension}) and (\ref{curvatureenergy}) that a non-standard choice for  $\Omega_{vac}$ doesn't change the numerical values of  $\alpha$ and  $\gamma$ for given $R$, $T$ and $\mu$.

Our results show that the surface tension is in the range of $\alpha \sim 145-165$ MeV/fm$^2$ and the curvature energy is in the range of $\gamma \sim 95-110$ MeV/fm.  
{The large values of the surface and curvature energies are due to the linear term in the expression for $\omega(x,y)$ in Eq. (\ref{omeguinha}), which is not present in the thermodynamic potential of e.g.  the MIT bag model.}
For a given $\mu$, the surface tension is an increasing function of $R$. The curvature energy behaves differently at constant $\mu$: for very small radii ($\sim 5 - 10$ fm) it increases with $R$, but in the range from $10 ~ \mathrm{fm}$ to $\infty $  it is a decreasing function of $R$.
Of course, both the surface and curvature contributions to the free energy per unit volume $\Omega_{MRE}/V$ {{tend}} to zero as $R \rightarrow \infty$, as can be checked from Eq. (\ref{eq17}). 
{The different behavior with $R$ for $\alpha$ and $\gamma$ is a consequence of the functional form of the integrand together with the R-dependence of the IR cutoff.
The integral for the surface tension has a fixed upper limit (the ultraviolet cutoff $\Lambda$), 
but the lower limit ($\Lambda_{IR}$) decreases with R. Then, as $R$ increases, 
$\Lambda_{IR}$ decreases, the area under the curve increases and as a result, 
the surface tension increases.
For the curvature energy the effect is different. The integrand	is positive for large $k$	but negative for small $k$. For small $R$, $\Lambda_{IR}$ falls in the positive region of the integrand. Thus, as $\Lambda_{IR}$ decreases, the positive area under the curve increases, and the curvature energy increases. For larger $R$, $\Lambda_{IR}$ falls in the negative regionof the integrand. Consequently,	as $\Lambda_{IR}$ decreases, the negative part of the area under the curve increases, and the curvature energy decreases.}
The results in Fig. \ref{fig1} show that the temperature dependence of $\alpha$ and $\gamma$ is very weak for the values of $T$ that are relevant for protoneutron stars and cold neutron stars.
Note also that these results are of the same order of the obtained within the NJL model  for just deconfined quark matter \textit{out of chemical equilibrium} under weak interactions \cite{Lugones2011,doCarmo2013}.

{The present equation of state can support a two solar mass neutron star such as the pulsars PSR J1614-2230 \citep{Demorest2010} and  PSR J0348-0432 \citep{Antoniadis2013} if a sufficiently stiff hadronic equation of state is employed for the outer layers of the star (see for example Fig 4 of \cite{Lenzi2012}). Adding a vector interaction to the NJL model, it is possible to stiffen the EOS and obtain larger stellar masses. The vector term affects the surface tension in a non-trivial way. The chemical potentials gain an extra term $\mu_{u,d,s} - 4 g_v \langle \psi^\dagger \psi \rangle_{u,d,s}$ that shift chemical equilibrium and the pressure gains a term proportional to the square of the density. The combined effect is difficult to estimate without a full calculation that will be addressed in future work. According to recent work \cite{Benghi2013a}, within a geometric approach to the surface tension evaluation, the surface tension can be lowered by the presence of a repulsive vector term and for magnetized quark matter the value will be further lowered \cite{Benghi2013b}. However, within the MRE formalism the behavior may be different, and deserves further study.} 

The large values of $\alpha$ and $\gamma$ have strong consequences for the physics of neutron star interiors, because the energy cost of forming quark drops within the mixed phase of hybrid stars would be very large.  According to \cite{Voskresensky2003}, beyond a limiting value of $\alpha \approx 65$ MeV/fm$^2$ the structure of the mixed phase becomes mechanically unstable  and local charge neutrality is recovered. Therefore, our results indicate that the hadron-quark interphase within a hybrid star should be a sharp discontinuity.

The consequences of the large values of $\alpha$ and $\gamma$ for the triggering of the deconfinement transition in neutron and protoneutron stars have been studied in \cite{Lugones2011,doCarmo2013}. The main difference with the present study of the quark-hadron interphase is the condition of chemical equilibrium. The nucleation (deconfinement) of the first quark matter drop that triggers the conversion of the core of a hadronic star is driven by strong interactions and consequently it happens out of chemical equilibrium  under weak interactions. 
As shown in Ref. \cite{Lugones2011}  the nucleation of quark matter is possible during the protoneutron star phase even for large values of the surface tension, because large drops (with a size of hundreds of fm) may have a huge nucleation rate. These large drops are charge neutral because flavor is conserved during the deconfinement transition \cite{Lugones2011} and therefore they can be considerably larger than the Debye screening length $\lambda_D$ of the stellar plasma which is typically $5-10$ fm. Since these drops can be very large ($R \gtrsim 200$ fm \cite{Lugones2011,doCarmo2013}), surface and curvature effects tend to vanish. This is not the case for droplets of quark matter in the hypothetical mixed phase of a hybrid star. Since they are electrically charged their size cannot exceed  $\sim \lambda_D \sim \mathrm{few \; fm}$ and therefore  surface and curvature have a significant energy cost, inhibiting the formation of the mixed phase.

\end{document}